\documentclass[twocolumn]{aastex631} 

\usepackage{graphics,epsf}
\usepackage[utf8]{inputenc}
\usepackage{amsmath}                
\usepackage{amsfonts}               
\usepackage{amssymb}                
\usepackage{epsfig}                 
\usepackage{graphicx}               
\usepackage{float}
\usepackage{color}
\usepackage{multirow}               
\usepackage{hyperref}

\hypersetup{
    colorlinks=true,
    linkcolor=red,   
    urlcolor=cyan}

\usepackage[colorinlistoftodos]{todonotes}

\newcommand{\s}{{~\rm s}}

\newcommand{\g}{{~\rm g}}
\newcommand{\G}{{~\rm G}}
\newcommand{\K}{{~\rm K}}

\begin{document}

\title{Identifying jittering-jet-shaped ejecta in the Cygnus Loop supernova remnant}
\date{July 2024}

\author[0000-0002-9444-9460]{Dmitry Shishkin}
\affiliation{Department of Physics, Technion, Haifa, 3200003, Israel; s.dmitry@campus.technion.ac.il; soker@physics.technion.ac.il}

\author{Roy Kaye}
\affiliation{Columbia University, New York, NY 10027, USA; rck2146@columbia.edu}
\affiliation{Department of Physics, Technion, Haifa, 3200003, Israel; s.dmitry@campus.technion.ac.il; soker@physics.technion.ac.il}

\author[0000-0003-0375-8987]{Noam Soker}
\affiliation{Department of Physics, Technion, Haifa, 3200003, Israel; s.dmitry@campus.technion.ac.il; soker@physics.technion.ac.il}

\begin{abstract}
Analyzing images of the Cygnus Loop, a core-collapse supernova (CCSN) remnant, in different emission bands, we identify a point-symmetrical morphology composed of three symmetry axes that we attribute to shaping by three pairs of jets. The main jet axis has an elongated S shape, appearing as a faint narrow zone in visible and UV. We term it the S-shaped hose, and the structure of three symmetry lines, the point-symmetric wind rose. The two other lines connect a protrusion (an ear or a bulge) with a hole on the opposite side of the center (a nozzle or a cavity), structures that we identify in the X-ray, UV, visible, IR, and/or radio images. There is a well-known blowout at the southern end of the S-shaped hose, and we identify a possible opposite blowout at the northern end of the S-shaped hose. The point-symmetrical morphology of the Cygnus Loop is according to the expectation of the jittering jets explosion mechanism (JJEM) of CCSNe, where several to few tens of pairs of jets with stochastically varying directions explode the star. The three pairs of jets that shaped the wind-rose structure of the Cygnus Loop are the last energetic pairs of this series of jets. Our study further supports the JJEM as the main explosion mechanism of CCSNe.  
\end{abstract}

\section{Introduction}
\label{sec:Introduction}

Identifying point-symmetric morphologies in several new core-collapse supernova remnants (CCSNRs) in 2024 constitutes a breakthrough in our understanding of the explosion mechanism of CCSNe. By 2023, studies identified point-symmetric morphologies and attributed them to the exploding jets in only two CCSNRs (see Table 1 in \citealt{Soker2023SNRclass}): the Vela CCSNR (for a newer study of Vela's point symmetry see \citealt{SokerShishkin2024}) and the identification of point symmetry \citep{Soker2022SNR0540} in the Doppler velocity maps of SNR 0540-69.3. 
New observations, like with JWST and eRosita, motivated a deeper search for point symmetry also in older observations, adding more CCSNRs with claimed jet-shaped point-symmetrical morphologies, CTB~1 \citep{BearSoker2023RNAAS}, 
N63A \citep{Soker2024CounterJet}, SN 1987A \citep{Soker2024NA1987A, Soker2024key},  G321.3–3.9 \citep{Soker2024CF, ShishkinSoker2024},  G107.7-5.1 \citep{Soker2024CF}, and Cassiopeia A \citep{BearSoker2024}. 
Comparison of these point-symmetric morphologies to jet-shaped point-symmetric structures of bubbles, lobes, and clumps in cooling flow clusters \citep{Soker2024CF} and planetary nebulae \citep{Soker2024PNSN} has solidified the claim that jets shaped the similar morphologies in CCSNRs. 
The typical large volume fraction of the point-symmetric structure out of the total CCSNR volume indicates that the energy of the shaping jets is comparable to the explosion energy of the CCSN, and the primary shaping can't be by post-explosion jets \citep{SokerShishkin2024}. Namely, jets exploded the CCSN progenitors of these CCSNRs. 

The shaping of CCSNRs with energetic jittering jets is the primary property of the jittering jets explosion mechanism (JJEM) of CCSNe. In the JJEM, the newly born neutron star (NS) (or black hole, if the NS collapses into a black hole) lunches pairs of jets as it accretes collapsing-core material with stochastic angular momentum. The stochastic angular momentum results from the stochastic convective motion fluctuations in the pre-collapse core (e.g., \citealt{ShishkinSoker2021, ShishkinSoker2023, WangShishkinSoker2024}) that are amplified by instabilities above the newly born NS. Post-explosion jets, interaction with the circumstellar material (CSM), or instabilities that develop in the explosion process cannot account for the observed properties of point-symmetric morphologies \citep{SokerShishkin2024}. All these processes, as well as NS natal-kick, neutrino heating \citep{Soker2022nu}, and the nucleosynthesis interaction of neutrinos with the ejecta, occur in the JJEM. However, only pairs of jittering jets (some with precession) can account for the point-symmetric morphologies \citep{SokerShishkin2024}. According to the JJEM, in rare cases of a rapidly rotating pre-collapse core, the newly born NS launches the exploding jets along a fixed axis. These jets do not remove much mass from the equatorial plane, and as a result, the NS continues to accrete mass and collapses into a black hole (e.g., \citealt{Soker2023gap}). A large amount of accreted mass can result in a super-energetic CCSN, with a small jittering around a fixed axis (e.g., \citealt{Gilkisetal2016}), and there are no failed supernovae in the JJEM. 

Recent studies of the alternative neutrino-driven (delayed-neutrino) mechanism focus on hydrodynamical and magneto-hydrodynamical simulations of the collapsing core aiming at reaching robust explosion (e.g., \citealt{Andresenetal2024, Burrowsetal2024, JankaKresse2024, Muler2024, Mulleretal2024, Nakamuraetal2024, vanBaaletal2024, WangBurrows2024, Huangetal2024}). The magneto-rotational explosion mechanism that requires a rapidly-rotating pre-collapse core, hence rare (e.g., \citealt{Shibagakietal2024, ZhaMullerPowell2024}), is considered part of the neutrino-driven mechanism as it still attributes most CCSNe to the neutrino-driven mechanism. However, none of these studies confronted the last-year claim that only pairs of energetic jets, which implies the exploding jets, can account for the newly identified point-symmetric CCSNR morphologies. As such, point symmetry is the most prominent property distinguishing between the two alternative CCSN explosion mechanisms. Its presence strongly supports the JJEM and poses a severe challenge, or even rules out, the delayed neutrino explosion mechanism. 

The ability of point-symmetric CCSNR morphologies to determine between the two explosion mechanisms motivates us to search for point-symmetric morphological features in as many CCSNRs as possible. The significant variations between CCSNRs in age, e.g., the young age of SN 1987A versus the old Cygnus Loop, in the properties of the surrounding interstellar medium (ISM), in explosion morphologies, and the variations in the components that appear in different wavelength and emission bands within individual CCSNRs, require a separate study of each CCSNR. This study identifies a point-symmetric morphology, which includes a precessing main jet axis, in the Cygnus Loop. In Section \ref{sec:MorphologicalFeatures} we describe the relevant morpholoical features, and in Section \ref{sec:PointSymmetry} we identify the point-symmetric morphology. Because the tool of point-symmetry identification is not standard in studying CCSNRs, unlike in planetary nebulae where it is prevalent (e.g., \citealt{Sahaietal2011, Clairmontetal2022, Danehkar2022, MoragaBaezetal2023, Mirandaetal2024}), we first list the properties we expect to find in point-symmetric CCSNRs (Section \ref{sec:Prescription}). We summarize in Section \ref{sec:Summary}.

\section{The point-symmetric prescription}
\label{sec:Prescription}

A point-symmetric morphology includes two or more pairs of similar morphological structures on opposite sides of the center, where the two pairs do not share the same symmetry axis.  In CCSNRs, we attribute these to pairs of jets. 
Such pairs might include bubbles (structures with bright rims around them), lobes (bubbles with partial rims), ears (bubbles with decreasing cross sections with distance from the central nebula), clumps, and filaments.  

This list of properties is crucial to identifying point-symmetric morphologies, the \textit{point-symmetric prescription}. 
\begin{enumerate}
\item \textbf{Some two pair's jets are unequal.} In the JJEM, the lifetime of an intermittent accretion disk that launches pairs of jets is typically $\simeq 0.01 -0.3 \s$ (e.g., \citealt{Soker2024key}) not much longer, or even shorter, than the relaxation time of the accretion disk. Therefore, the accretion disk formed by stochastic fluctuations might have two unequal sides that, in turn, launch two jets unequal in their power and opening angle \citep{Soker2024CounterJet}. Therefore, a pair's two opposite structural features might sometimes be at different distances from the center, sizes, and shapes.  
\item \textbf{Rim-nozzle asymmetry.} A prominent asymmetrical structure (also in planetary nebulae) is one where, on one side, the front of a jet-inflated bubble is closed and appears as a rim, and on the other side, the jet has broken out, leaving a nozzle (opening) in the front boundary of a jet-inflated bubble. Prominent CCSNRs with this rim-nozzle asymmetry are SN 1987A, SNR G309.2-00.6, and  SNR G107.7-5.1 \citep{Soker2024PNSN}  
 \item \textbf{A main jet axis.}  As the accretion rate decreases towards the end of the explosion process, the angular momentum fluctuations also decrease, allowing the last pair or two of jets to live long and leave a prominent axially-symmetric signature (precession is possible) on the CCSNR \citep{Soker2024key}. This is the main jet axis. Point-symmetric CCSNRs with main jet axis include SN 1987A (its `keyhole'; \citealt{Soker2024key}), the Vela SNR \citep{SokerShishkin2024},  and SNR 0540-69.3 \citep{Soker2022SNR0540, Soker2024CF}. In this study, we add the Cygnus Loop to this list.  
 \item \textbf{Ejecta composition.} Point-symmetric structures that are rich in elemental abundance from the inner core, like Fe, Si, S, O, Ne, or Mg, show that they were ejected by jets during the explosion process and not after the explosion (e.g., \citealt{SokerShishkin2024}.   
 \item \textbf{NS natal kick.} The NS natal kick in the JJEM is similar to the neutrino-driven explosion mechanism but avoids small angles with the main-jet axis (e.g., \citealt{BearSoker2023RNAAS}). The large angle of the NS kick direction to the main jet axis implies that the NS needs not to be along the symmetry axis of the main jet axis, nor at the point where different symmetry axes of the point-symmetrical morphology cross each other (e.g., as in Cassiopeia A; \citealt{BearSoker2024}). 
 \item \textbf{Interaction with ISM and the CSM.} The ISM, in particular in old CCSNRs, can substantially change the structure in the outer ejecta (e.g., \citealt{Wuetal2019, YanLuetal2020, LuYanetal2021, MeyerMelianietal2024, Sofue2024}), as is the case with the Cygnus Loop (Section \ref{sec:MorphologicalFeatures}). Interacting with the ISM and circumstellar material (CSM) sends a reverse shock into the inner ejecta. Therefore, the ISM and CSM might also influence the appearance of the inner ejecta. 
 \item \textbf{Point symmetry from the CSM}. 
 The CSM influences the late morphologies of CCSNR (e.g., \citealt{Chiotellisetal2021, ChiotellisZapartasMeyer2024, Meyeretal2022, Velazquezetal2023}; \citealt{MeyerDetal2024} study the Cygnus Loop). Some point-symmetric morphological features might result from the CSM if it has such a morphology. A prominent example is the three CSM rings of SN 1987A that will imprint their structure on the SN 1987A remnant in a few hundred years. This process, though, cannot explain different elemental distributions nor point-symmetry in the inner ejecta.    
 \item \textbf{Instabilities.} The JJEM involves strong instabilities. Numerical simulations show vigorous instabilities in neutrino-driven explosions (e.g., \citealt{Wongwathanaratetal2015}). Similar instabilities are expected in the JJEM. The instabilities are stochastic and cannot form point-symmetrical structures; they can smear such structures. The instabilities might increase the asymmetry between the pair's two opposite structural features.  
 \item \textbf{Nickel bubbles.} The ejecta expansion is not purely homologous because its sound speed might sometimes be non-negligible. It is hot immediately after the explosion, and then heat deposition by radioactive decay heats the ejecta, and later, the reverse shock (which might have a complicated geometer, e.g.,  \citealt{HwangLaming2012}) also heats the ejecta.  
The heating by local radioactive decay of nickel and then cobalt are termed nickel bubbles. They, for example, can form rings, bubbles, and knots, as observed in Cassiopeia A (e.g., \citealt{HammellFesen2008, MilisavljevicFesen2013, FesenMilisavljevic2016}). The hot ejecta expands in all directions, a flow that smears the point-symmetrical features.
 \end{enumerate}

Considering all these processes and assuming that jittering jets explode most CCSNe (the JJEM), we set the goal of revealing the point-symmetric morphology of the Cygnus Loop. 

\section{The relevant morphology features}
\label{sec:MorphologicalFeatures}

This section describes the relevant morphological features for identifying the point-symmetrical morphology (Section \ref{sec:PointSymmetry}). 
We use several surveys and data sources ranging from radio to x-ray, as we detail in figure captions. Some files used for image creation were retrieved using the \textit{SkyView} tool.

\begin{figure*}[b!]
\begin{center}
\includegraphics[trim=2cm 3cm 0.5cm 5cm,width=0.96\textwidth]{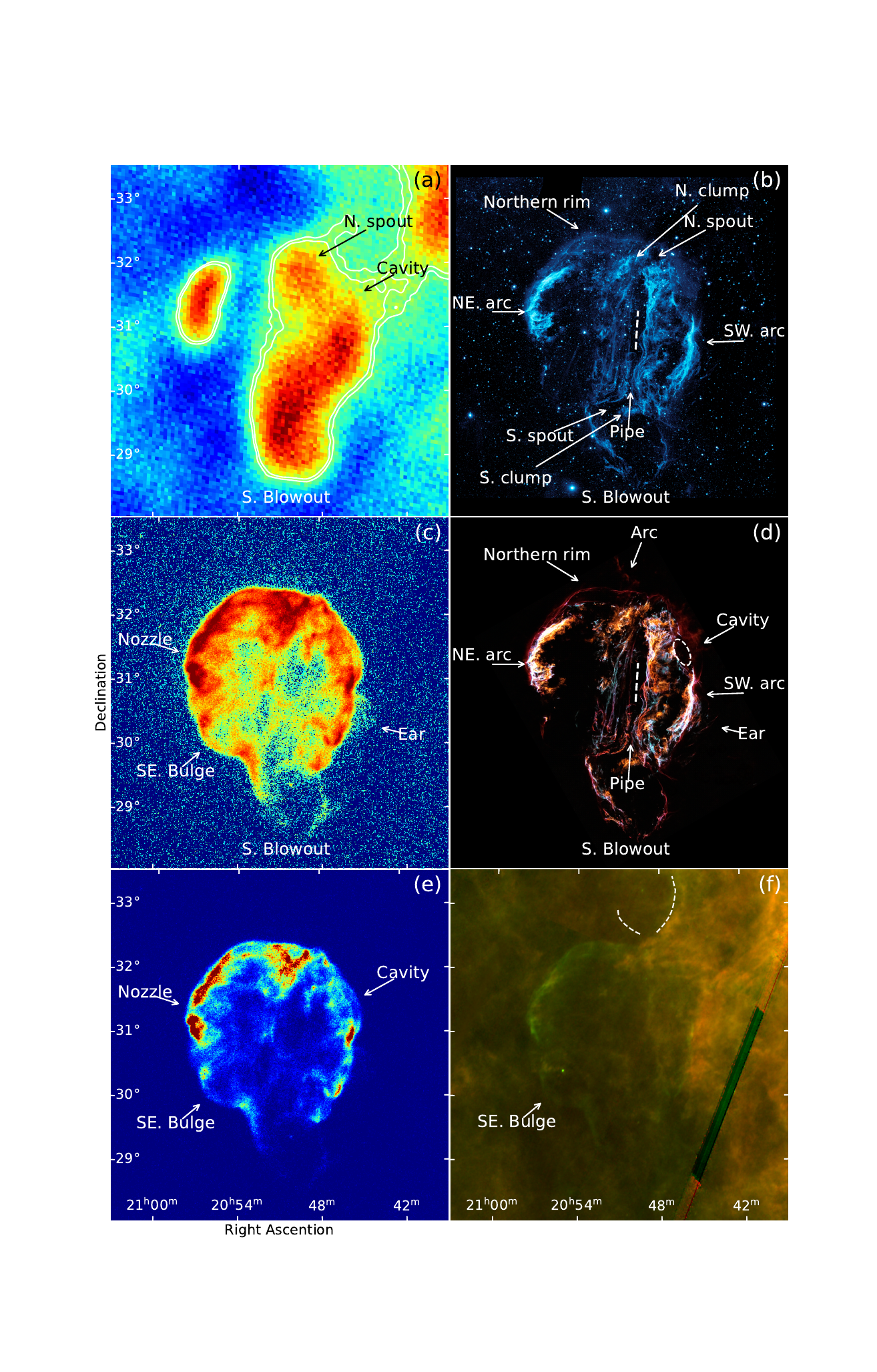} 
\caption{See next page}
\label{fig:6panels}
\end{center}
\end{figure*}

We present images of the Cygnus Loop in the different bands in Figure \ref{fig:6panels} and mark the different morphological features we elaborate on. 
\addtocounter{figure}{-1}
\begin{figure*} [t!]
\caption{(Previous page.) The Cygnus Loop SNR in several wavelengths. (a) Planck 30GHz I (\citealt{PlanckLFI2018}; linear scale, truncated at $1.1mK$); (b) GALEX UV (\citealt{GALEX2007}; $175-280 \rm{nm}$; Image Credit: NASA/JPL-Caltech); (c,e) ROSAT X-ray count image (\citealt{ROSAT1993}; Soft band: $0.1-0.4 keV$; $0.1-50 \rm{counts}$ range) in log (panel c) and linear (panel e) scale; (d) SHO (visible) image \citep{Raymondetal2023}; (f) AKARI IR ($140$ and $90 \rm{\mu m}$, $\rm MJy/sr$).
We identify the morphological features used to identify the point-symmetric morphology (Section \ref{sec:PointSymmetry}). The short dashed-white line in panels b and d marks a segment of the pipe, which extends from the north spout to the south spout and connects them. The cavity is the dark area in the visible band that the dashed-white ellipse marks on panel d. The dashed curved lines in panel f depict the arc seen in panel d (left) and the curved edge of the dust cloud to the north of the Cygnus Loop (right). Contour lines in panel a are of a kernel smoothed image, at $\sim40\%$ and $\sim50\%$ levels. Marking locations are consistent between different panels.}
\end{figure*}

\textit{The southwest (SW) and northeast (NE) bright arcs. } The UV (panel b) and visible (panel d) bands show two opposite bright arcs on the boundary of the main SNR shell, as we mark on these panels. The X-ray image of panels c and e also reveals these arcs. The SW arc is almost flat, while the NE arc has a small protrusion with a nozzle in its center.  

\textit{The nozzle in the northeast (NE) bright arc.} Panel e of Figure \ref{fig:6panels} reveals an opening in the northeast (NE) bright arc (between two red tangential segments; also on panel c). A small break in the arc is seen in the GALEX image (panel b) and in the AKARI image (panel f). 

\textit{The ear on the southwest arc.} A log-scaled X-ray image in panel c of Figure \ref{fig:6panels} reveals an `ear' on the west-southwest protruding from the bright southwest arc. An `ear' is a protrusion from the main body of the ejecta/nebula, with a base smaller than the main ejecta/nebula, a monotonically decreasing cross-section with distance from the center, and somewhat different properties than the main ejecta/nebulae, e.g., fainter in some bands and/or different structural small-scale features. The ear appears also in the UV (panel b) and visible (panel d).  
Following several studies of ears in CCSNRs (e.g., \citealt{GrichenerSoker2017, Bearetal2017} and papers identifying point-symmetry in other CCSNRs,  see Section \ref{sec:Introduction}) that consider jets to shape the ears in CCSNRs, we take the ear in the west-southwest of the Cygnus Loop to be an imprint of a jet. 

There is a possible alternative explanation for the ear. We note that the edge of the main ejecta at the base of the ear and its sides form a more or less flat boundary to the main nebula on the southwest. Such an ear morphology protruding from a flat boundary can result from Rayleigh-Taylor instability when a low-density ambient gas decelerates a denser ejecta. \cite{Martinetal2002} discovered such a structure in the planetary nebula NGC 40 and termed the protrusion `bump.' The bump looks like an ear and is protruding in the direction of motion of NGC 40 through the ISM, as expected for Rayleigh-Taylor instability. If this explains the ear in the southeast of the Cygnus Loop, then that side of the ejecta was substantially decelerated. This implies that the center of the ejecta (explosion site) is much closer to the southwest boundary than the northeast boundary. This is also compatible with the explosion center we suggest in this study (Section \ref{sec:PointSymmetry}). 

Because the ear is on the opposite side of the nozzle in the northeast arc, we will assume that a jet inflated the ear in the southwest. However, we cannot rule out the possibility that the ear results from Rayleigh-Taylor instability in the ejecta-cloud interaction.

\textit{The southeast (SE) bulge.} The southeast bulge is a small protrusion from the main SNR shell, appearing as a faint rim in the X-ray (panel c) and IR image (panel f). The curvature of the rim of the bulge is higher (smaller curvature radius) than the main SNR shell (it might be alternatively classified as a shallow ear).

\textit{The thin rim the north.}  The UV (panel b) and visible (panel d) show a thin rim on the north of the SNR. Several filaments, studied by, e.g., \cite{Vucetic_etal_2023}, compose the northern rim.  

\textit{The northwest cavity.} The northwest cavity is an oval-shaped dark zone in the northwest boundary of the main SNR shell most prominent in the visible, as we mark on panel d of Figure \ref{fig:6panels}. The cavity is also slightly apparent in the UV (panel b). A structure perpendicular to the cavity and extending from the city outwards to the northwest is seen in both the visible (panel d; $\rm H\alpha$, red) and radio (panel a; $1 \rm cm$).

\textit{The group of thin north-south filaments and the `pipe.'} The radiation-emitting filaments in the Cygnus Loop are thin sheets of post-shock gas tangent to the line of sight \citep{Raymondetal1988}. We now refer to the thin filaments along the north-south direction that the UV (panel b in Figure \ref{fig:6panels}) and visible (panel d) bands clearly show. In between these filaments is a dark north-south elongated zone, the `pipe.' If a shock front is close to being parallel to the plane of the sky, the surface brightness is too low to be detected. We assume that there is a shock gas around the `pipe' (later, we take it to be the main-jet axis), but the sheets in front and behind the `pipe' are close to parallel to the plane of the sky, hence below the detection limit. 

\textit{The pipe and its two opposite spouts.} The `pipe' we defined above is bounded by two opposite bent segments that we term spouts, the north and the south spout. A bright clump is outside each bending. 

\textit{The blowout on the south.} A large blowout in the south of the Cygnus loop appears in different bands. The visible and UV bands (panels d and b) reveal its full structure. It looks as if the southern spout has blown this blowout. The blowout volume is filled with material that emits in the radio (panel a of Figure \ref{fig:6panels}). 

\begin{figure*}[t]
\begin{center}
\includegraphics[trim=2cm 2cm 0.5cm 3cm,width=\textwidth]{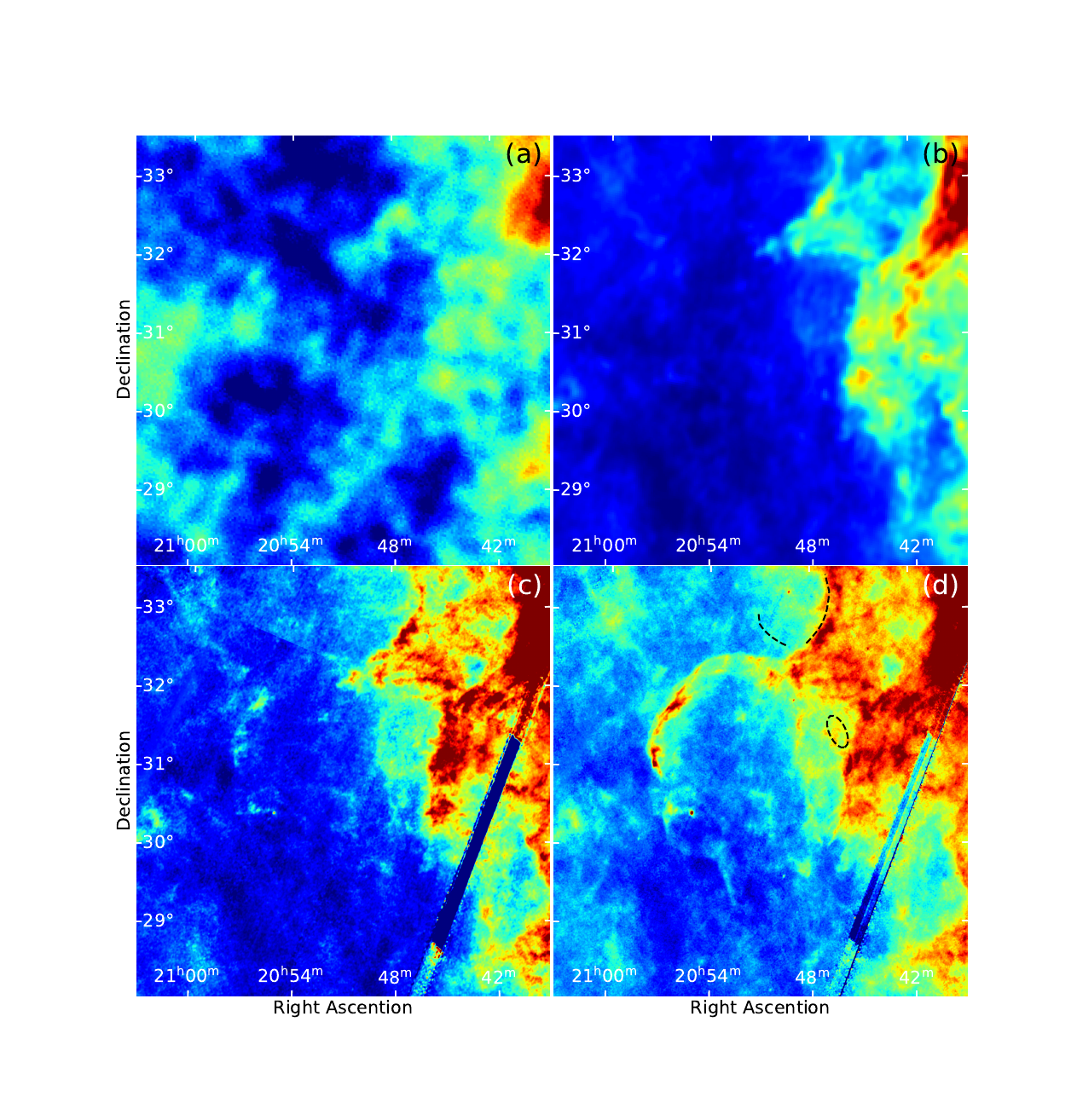} 
\caption{Four panels at different IR emission wavelengths covering the same area of the panels in Figure~\ref{fig:6panels}. These images highlight the dust cloud surrounding the Cygnus Loop remnant. Panels (a) and (b) are Planck images 143GHz I ($\rm K$) and 857GHz I ($\rm MJy/sr$), respectively. Panels (c) and (d) are AKARI images at $140\mu m$ and $90\mu m$ ($\rm MJy/sr$), respectively. On panel (d), we mark the curved edge of the dust cloud to the north of the Cygnus Loop and the arc location from Figure \ref{fig:6panels}. We also denote the location of the cavity from panel d of Figure \ref{fig:6panels} with a dashed ellipse for reference.}
\label{fig:dustIR}
\end{center}
\end{figure*}

\textit{The Arc.} A faint and curved arc is seen only in panel d of Figure \ref{fig:6panels}. It is connected to the northern part of the Cygnus Loop. We draw this arc on panel f of Figure \ref{fig:6panels}, where a dust cloud edge is seen in the IR, also marked with a dashed line. We speculate there might be a counterpart to the southern blowout, bound by the arc (on the east) and the cloud (on the west). We also mark these two features in panel d of Figure \ref{fig:dustIR}.

\textit{The clouds.} In Figure \ref{fig:dustIR} we present the Cygnus Loop and its surroundings in the IR at four wavelengths. Observable in the IR is a dense dust cloud, mainly located west to the remnant. A small ``cloudy'' structure also appears on the eastern front, coinciding with the northeastern bright arc. 
%

\section{The point-symmetric morphology of the Cygnus Loop }
\label{sec:PointSymmetry}

Based on the morphological features that we identified in Section \ref{sec:MorphologicalFeatures} and marked on Figures \ref{fig:6panels} and \ref{fig:dustIR}, we identify here a point-symmetric structure in the Cygnus Loop. In doing so, one should consider the different processes that cause deviation from perfect point symmetry, including the NS natal kick, the interaction of the ejecta with the ISM and the CSM, and large-scale instabilities, as we discussed in Section \ref{sec:Prescription}. 

\begin{figure*}[t]
\begin{center}
\includegraphics[trim=2cm 2cm 0.5cm 3cm,width=\textwidth]{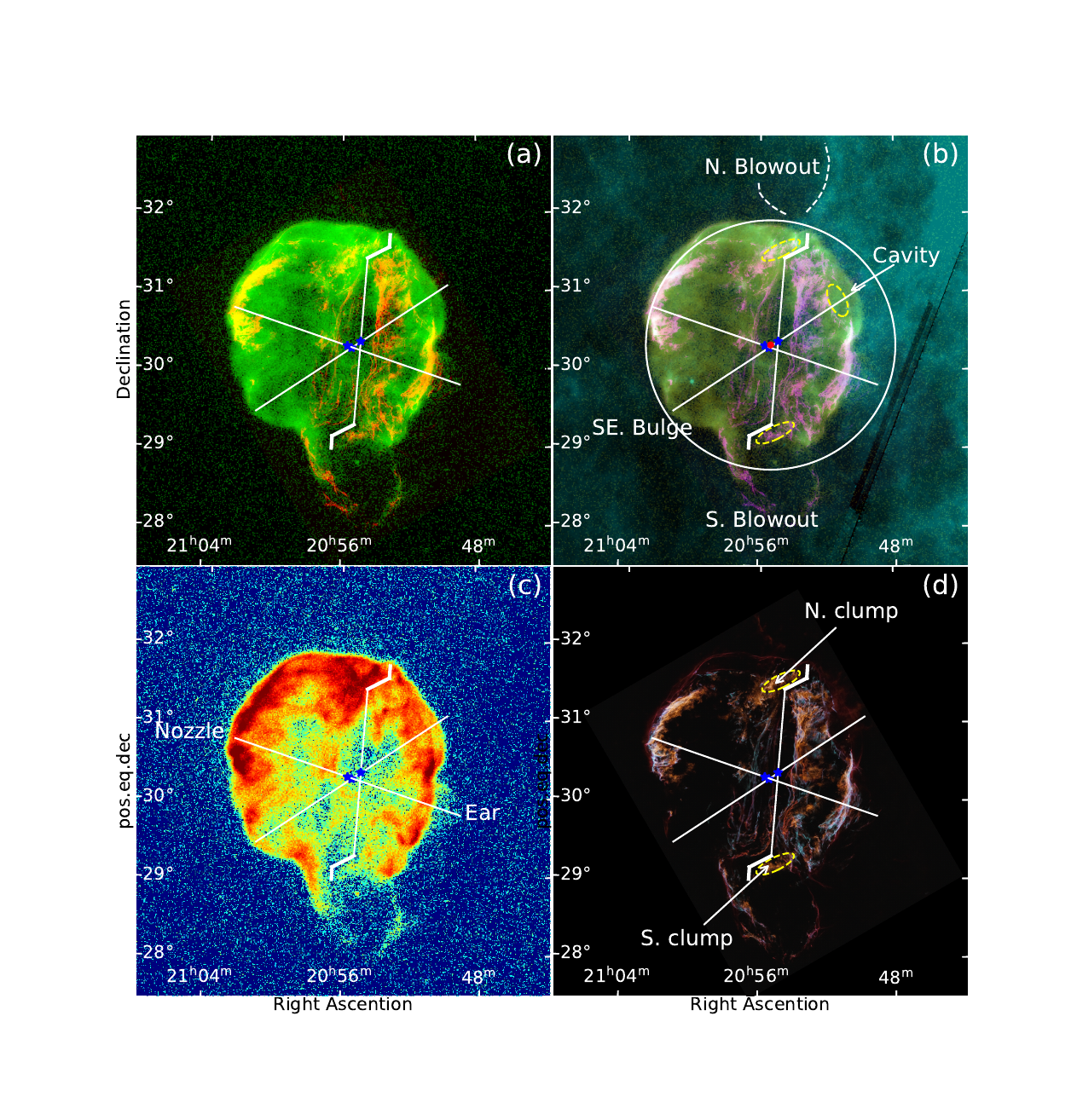} 
\caption{Four annotated panels pointing out the point-symmetrical structure we identify in the Cygnus Loop. (a) Red depicts the visible image (panel d of Figure \ref{fig:6panels} and panel d here) and green depicts log scale x-ray images (panel e in figure \ref{fig:6panels} and panel c here); (b) Magenta represents visible (panel d of Figure \ref{fig:6panels} and panel d here), the green represents log scale x-ray (panel c in figure \ref{fig:6panels} and panel c here), and the teal represents the AKARI $90 \mu m$ image (panel d in Figure \ref{fig:dustIR}); (c) Log scale x-ray; (d) visible as panel d in Fig.~\ref{fig:6panels}. The pipe and the two spouts (see Figure \ref{fig:6panels}) form the `hose', i.e., the long and narrow faint zone extending north-south. The structure of the hose with the two clumps (dashed yellow ellipses) has a $180^\circ$ symmetry around its center (marked by a blue star at its center). The two other lines connect two opposite structures, as indicated, with a blue star marking the center of each line. The three symmetry lines form the point-symmetric \textit{wind-rose} of the Cygnus Loop. We attribute this shaping to at least three pairs of opposite jets.  
A red circle in panel b denotes the average location of the three centers (in the plane of the sky). We also denote a possible northern blowout with two curved lines in panel b, bounded by the visible arc (panel d of Figure~\ref{fig:6panels}) and the curved edge of the dust cloud also seen in Figure~\ref{fig:dustIR}. In panel b we also mark a circle with a radius matching the eastern edge of the remnant whose center is the average of the \textit{wind rose} (red circle).}
\label{fig:pointSym}
\end{center}
\end{figure*}

We identified an S-shaped structure, the \textit{S-shaped hose}, as most prominently appears in the visible image (panel d of figure \ref{fig:pointSym}). The S-shaped hose includes the pipe as its central segment, the two spouts at the ends of the pipe, and a short extension at the end of each spout. We mark the S-shaped hose with the five-segment solid-white line on the different panels of Figure \ref{fig:pointSym}. The star on the pipe is the center of the S-shaped hose, which has a $180^\circ$ rotation symmetry around that center. The S-shaped hose is adjacent to the radio filament that \cite{Leahy1999} termed the ``Serpent.'' In addition to the 5 segments of the hose, there is a prominent large southern blowout attached to the southern part of the hose. As we discussed in Section \ref{sec:MorphologicalFeatures}, we identify a possible blowout in the north as well. We mark these on panel (b) of Figure \ref{fig:pointSym}. The two blowouts are different in shape and size but have a symmetric location with respect to the hose. The two clumps (see panel b of Figure \ref{fig:6panels}) also have symmetrical locations with respect to the hose. We conclude that the hose presents a robust point-symmetrical structure in the Cygnus loop. It might have been shaped by a precessing pair of jets. 

We identified two other pairs of structural features that, together with the hose, form the \textit{point-symmetric wind-rose} of the Cygnus Loop. In the first, we connect the Nozzle (panels c and e of Figure \ref{fig:6panels}) and the ear (panels c and d) of Figure \ref{fig:6panels}). We mark this pair with a thick line in Figure \ref{fig:pointSym}, going from northeast to southwest; the star on this line marks its center. We also identify a pair by connecting the bulge (SE. Bulge in panels c, e and f of Figure \ref{fig:6panels}) with the cavity (panel d of Figure \ref{fig:6panels}); this is the less prominent pair in the point-symmetric wind-rose. 
The three lines of the wind rose are very close to crossing each other at the same point. The centers of the two lines are to the east (left) of the hose. This is expected because the large ISM cloud on the west slows down the western part of the Cygnus Loop. Other perturbations, like the NS kick, can further move the centers apart. 

We attribute the Cygnus Loop's point-symmetric structure to the three last energetic pairs of jets that exploded its progenitor in the frame of the JJEM. The pipe, the hose's central and longest part, is the Cygnus loop's \textit{main jet axis}. Several CCSNRs have a main jet axis, e.g., SN 1987a. A long-lived pair of jets in the final phase of the explosion process might form such a structure \citep{Soker2024key}.

The IR/radio images (Figure \ref{fig:dustIR}) show a large cloud on the entire western side of the Cygnus Loop. We consider that this cloud has substantially slowed down the expansion of the west side of the Cygnus loop. This is evident from the large-scale flat structure of the southwest arc (beside the ear). To further demonstrate this with the general point symmetric structure we identified, we draw a circle with its center near the cross points of the three lines of the point-symmetric wind-rose (panel b of Figure \ref{fig:pointSym}). We make the eastern part of this circle to fit the east boundary of the Cygnus Loop. We can see that the western side of the circle is much larger than the west boundary of the Cygnus Loop, beside the tip of the ear that touches the circle.

\section{Discussion and summary}
\label{sec:Summary}

Using some morphological features of the Cygnus Loop that we described in Section \ref{sec:MorphologicalFeatures} and marked in Figures  
\ref{fig:6panels} and \ref{fig:dustIR}, in Section \ref{sec:PointSymmetry}, we identified three symmetry axes that together form a \textit{point-symmetric wind-rose} that signify the point-symmetric morphology of the Cygnus Loop; we drew the wind rose in figure \ref{fig:pointSym}. The S-shaped hose is the five-segment line, extending from the possible blowout in the north to the clear blowout in the south. Its central straight line, the pipe (Figure \ref{fig:6panels}), is the main jet axis of the Cygnus Loop. We attribute the point-symmetric morphology to the last three energetic jets in a series of several pairs of jets to a few tens of pairs of jets that exploded the Cygnus Loop's progenitor in the JJEM frame.    

Our study adds to the rapidly growing observational support of the JJEM as the main, or even only, explosion mechanism of CCSNe. The alternative delayed-neutrino explosion mechanism cannot account for point symmetric morphologies (see discussion in section \ref{sec:Introduction}), nor can interaction with the CSM or ISM (e.g., \citealt{SokerShishkin2024}).  

\begin{figure*}[t]
\begin{center}
\includegraphics[trim=2cm 18cm 0.5cm 3cm,width=\textwidth]{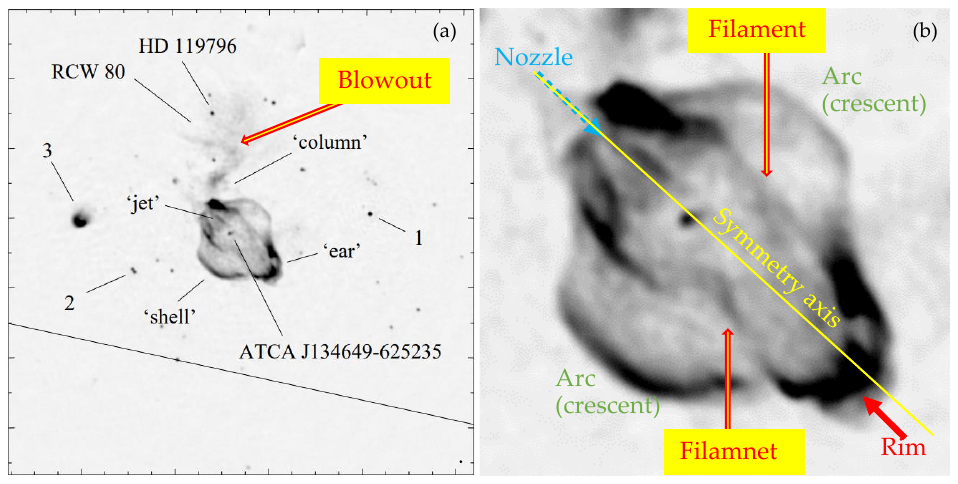} 
\caption{(a) A radio image of SNR G309.2–00.6 and marks as presented by \cite{Gaensleretal1998}, who already argued that jets shaped this CCSN. (b) The main CCSNR shell with labels from \cite{Soker2024PNSN} of the two arcs (`crescents’), the symmetry axis, and the rim-nozzle structure. We added the labelling of the blowout (panel a) and the filaments (panel b). }
\label{fig:SNRG309}
\end{center}
\end{figure*}

Most of the individual morphological features we use to identify the point symmetry of the Cygnus Loop are not unique to this CCSNR. Ears are very common in CCSNRs (e.g., \citealt{GrichenerSoker2017}), and so are nozzles and rims (e.g., \citealt{Soker2024PNSN}). In the Cygnus Loop one symmetry axis connects a nozzle and an ear. In many cases, rims are the bright outer front of an ear. 
Consider the SNR G309.2-00.6 (The stingray remnant; also: G309.2-0.6). \cite{Gaensleretal1998} studied and imaged SNR G309.2–0.6 in the radio band and suggested and discussed its shaping by jets. \cite{YuFang2018} simulated jet shaping and obtain the general morphology SNR G309.2–0.6, but not all details. \cite{YuFang2018} also estimated the energy of the two jets that inflated the two observed ears to be $\simeq 10-15\%$ of the explosion energy, while \cite{GrichenerSoker2017} estimated this energy to be $\simeq 7 \%$ of the explosion energy. SNR G309.2-0.6 has one symmetry axis with a rim-nozzle asymmetry. We present this CCSNR in Figure \ref{fig:SNRG309} because it has other similarities with the Cygnus Loop: It has a central zone defined by an elongated structure, on which we mark the two filaments, and two arcs on the outskirts to the side of the symmetry axis. SN 1987A, now a CCSNR 1987A, also has two arcs to the sides of its rim-nozzle elongated structure \citep{Soker2024PNSN}; \cite{Matsuuraetal2024} discovered the arcs and termed them crescents. On the end of the nozzle, there is a blowout in SNR G309.2-0.6. A nozzle is probably where a jet broke out from the main CCSN shell; such a jet can inflate a blowout. {{{{ \cite{YuFang2018} simulated the interaction of jets with ejecta and form ears. Their simulated jets have a half-opening angle of $10^\circ$, which seems too large to break from the ejecta. The narrow pipe we identify in the Cygnus Loop also suggests that forming a nozzle through which a jet can create a blowout requires narrow jets with a half-opening angle of no more than a few degrees. }}}}
Interaction with the CSM, as studied by, e.g., \cite{Gvaramadze1999} and \cite{Chiotellisetal2021}, cannot explain the inner filaments with outer arcs nor the blowout. {{{{ The numerical simulations we mention above present the density maps of the interaction. At the same time, the radio emission depends on the intensity of magnetic fields and the acceleration of electrons to high energies. Therefore, morphologies from simulations should be compared cautiously with radio maps.   }}}}
The comparison with SNR G309.2–0.6 strengthens our identification of the jet-shaped point symmetric morphology of the Cygnus Loop. 

\begin{figure}[t]
\begin{center}
\includegraphics[trim=5cm 1cm 2cm 0cm,width=0.5\textwidth]{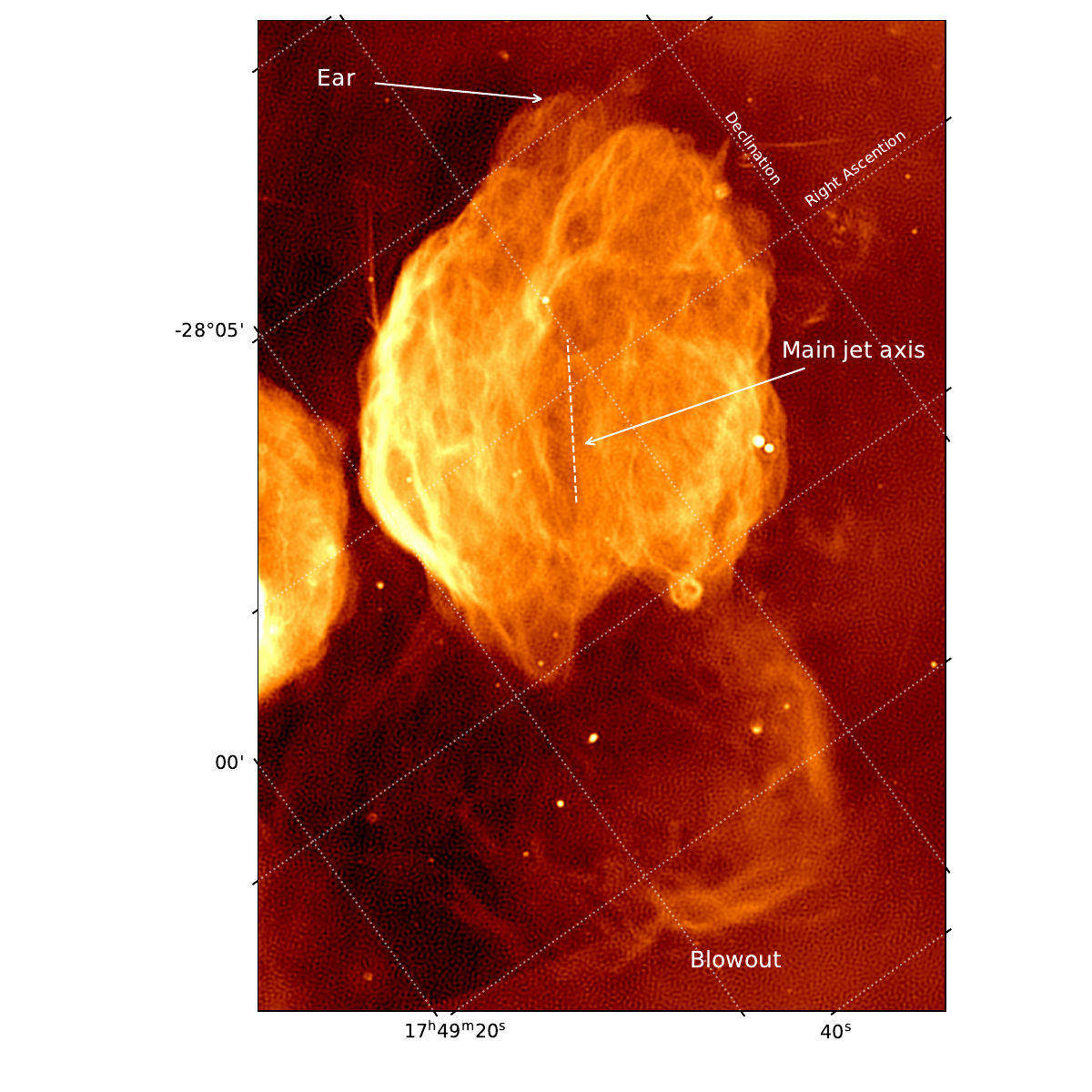} 
\caption{A small area of the full MeerKAT (radio emission) total-intensity galactic center mosaic \citep{HeywoodMeerKAT_G1.0-0.1_2022}. We added the labels and white dashed line on the image of the SNR G1.0-0.1: The blowout, the ear, and what we suggest to be the main jet axis. We argue this CCSNR was also shaped by pairs of jets, the last jets of the many that exploded this CCSNR in the frame of the JJEM. Grid lines are in dotted white lines and labels denoted at top right.}
\label{fig:SNRG1001}
\end{center}
\end{figure}

Another possible example of a blowout from a CCSNR is the SNR G1.0-0.1 at the galactic center. We present this CCSNR in Figure \ref{fig:SNRG1001}, based on data described in \cite{HeywoodMeerKAT_G1.0-0.1_2022}. This SNR deserves a separate study; we will not explore more details here. 

The short summary is that the point-symmetrical morphology we identified in the Cygnus Loop CCSNR, i.e., its point-symmetrical wind rose, is robust and in support of the JJEM of CCSNe. 

\section*{Acknowledgements}
We thank John Raymond for very helpful explanations and clarifications.
A grant from the Pazy Foundation supported this research.

We acknowledge the use of NASA's \textit{SkyView} facility (https://skyview.gsfc.nasa.gov) located at NASA Goddard Space Flight Center.

Some of our images in the IR are based on observations with AKARI, a JAXA project with the participation of ESA.
Other far-IR/radio images were extracted from Planck Public Data Release 3 (Planck Collaboration). 

For the visible data we use the SHO image produced by Min Xie \citep{Raymondetal2023}.

We use the Galaxy Evolution Explorer (GALEX) survey GR6/7 data for the near-UV images, image Credit: NASA/JPL-Caltech.

For our xray data, we have made use of the ROSAT Data Archive of the Max-Planck-Institut für extraterrestrische Physik (MPE) at Garching, Germany.

The MeerKAT telescope is operated by the South African Radio Astronomy Observatory, which is a facility of the National Research Foundation, an agency of the Department of Science and Innovation.

\end{document}